# CO-phylum：An Assembly-Free Phylogenomic Approach for Close Related Organisms

Huiguang Yi

State Key Laboratory of Genetic Engineering and MOE Key Laboratory of Contemporary Anthropology, School of Life Sciences and Institutes of Biomedical Sciences at Fudan University

## Abstract

Phylogenomic approaches developed thus far are either too time-consuming or lack a evolutionary basis. Moreover, no phylogenomic approach is capable of constructing a tree directly from unassembled raw sequencing data. A new phylogenomic method, CO-phylum, developed to alleviate these flaws. CO-phylum can generate a high-resolution and highly accurate tree using complete genome or unassembled sequencing data of close related organisms, in addition, CO-phylum distance is almost linear with p-distance.

## Introduction

One important field of study in biology is the evolutionary history and phylogenetic relationship between organisms. Darwin's creative explanations of the relationships between the extensive biological morphology data gave rise to the study of phylogenetics. With the development of molecular biology, phylogenetic studies shifted their focus from morphology to nucleotide or amino acid sequence data. Based on multiple sequence alignments of orthologous genes, phylogeneticists could trace biological characteristics to the nucleotide or amino acid level. In prokaryotes, the multiple sequence alignments of 16S rRNA sequences has become a standard method in the phylogenetic analysis because 16S rRNA exists in almost all prokaryotic organisms, and it rarely undergoes horizontal gene transfer (HGT). However, 16S rRNA is highly conserved, which can make sequences comparison difficult in two close related organisms, thus the phylogenetic method based on 16S rRNA sequences is difficult to constructs a high-resolution tree. Currently, thousands of prokaryotic genomes are available, and with the ongoing development of next-generation sequencing (NGS) technology, phylogenetic analysis of dozens or even hundreds of organisms from a single genus could be possible. To match these technological advances, ultrahigh-resolution phylogenetic methods must be developed. Less conserved genes could be used to build a higher resolution phylogenetic tree; however, gene selection would be controversial. In different genera, different genes might be chosen, and it is would be especially hard for non-experts to make this selection. In addition, the potential issue of HGT may greatly influence the performance of phylogenetic analysis based on individual genes. Numerous studies have attempted to construct phylogenetic relationships based on whole genome data, termed phylogenomics. The phylogenomic methods proposed thus far could be roughly divided into alignment-based methods and alignment-free methods. The former includes methods such as supermatrix and supertree(1), gene content (2) and gene order (3) etc.. The later includes methods such as the 'average common substring' approach(4) and composition vector tree (CVtree) (5) etc.. In the supermatrix and supertree approaches, multiple-sequence alignments are employed for whole genome-wide orthologous genes. Usually, different gene alignments have unequal length and contain different sets of taxa units. The supermatrix approach concatenates all individual gene alignments, and the concatenated gene alignment is then used to deduce the phylogenetic tree. Supertree will first obtain an optimal tree from individual gene alignments and then combined the trees. Gene-content and gene-order approaches will first identify orthologous sequences in the paired genomes and then the gene-content approach constructs phylogenetic trees from 'distances' that represents the proportion of shared orthologous genes between genomes, gene-order approach construct phylogenetic trees by minimizing the number of breakpoints between genomes or by scoring the presence or absence of pairs of orthologous genes. Compare with supermatrix and supertree approaches, gene-content and gene-order approaches are more focused on structural variation but the sequence level nucleotide change. All of these alignment-based phylogenomic methods have the same flaw: they should identify genome-wide orthologous segments before construct a tree, this progress is very time-consuming and would be difficult to put into practice when meet huge amount of genomes. Alignment-free methods are quite different with the alignment-based methods, they don't need to identify the orthologous segments, instead, they usually calculate distances of pair wise organisms based on oligo-peptide word usage frequency (5,6) or information theory tools, such as Kolmogorov complexity (7,8), Lempel–Ziv complexity(9). Once all pair wise distances are calculated, a distances matrix would be generated, standard methods such as neighbor-joining (NJ) (10) then would be employed to construct the phylogenetic tree using the distances matrix. A approach proposed recently, the 'average common substring', is based on another information theory tool 'Kullback–Leibler relative entropy'(4) and the distance in average common substring approach in fact reflects the average length of the maximum common substring of the paired sequences. Composition vector tree (CVtree) (5), singular value decomposition (6) (SVD) and recent feature frequency profiles (FFPs) method (11) are similar approaches based on 'word frequency' where, CVtree for example, each proteome are represented as a vector of K-peptide (word) frequencies and the pair wise sequence distance is measured using the cosine of the angle between the corresponding vectors. These alignment-free phylogenomic methods are much faster than alignment-based phylogenomic methods because they circumvent the time-consuming process of genome-wide orthologous segments identifying. However, these alignment-free phylogenomic methods have themselves problems, for example, distances measured based on information theory or word usage frequency usually don't have a solid evolutionary basis and they hardly be linear with evolutionary time.

All the phylogenomic methods developed thus far are unable to utilize unassembled sequencing data. However, with the adoption of NGS, Reads generated from NGS are much shorter than traditional Sanger sequencing, which makes complete genomes assembly a more difficult to impossible task, especially when the genome contains large proportion of repetitive segments. This raise a question: could we complete the phylogenomic analysis without an assembled genome?

Here, we propose a new phylogenomic approach, CO-phylum, which not only has algorithm efficiency as high as other alignment-free approaches but also can take advantage of both complete genome and unassembled NGS data. Most importantly, the distance represented in CO-phylum is based on the theory of molecular evolution but information theory or 'word frequency', so long as the distance is not too large, it keep very well linear relationship with p-distance. In several genera we test thus far, CO-phylum was able to generate a high-resolution tree using either complete genome data or NGS data. The trees we constructed were highly concordant in both topology and branch length with those trees constructed using methods based on genome-wide alignments.

## Materials and Methods

1. Introduction to algorithm

CO-phylum is based on the 'Context-Object' idea in which an 'Object' is a nucleotide in the genome (or an acid amino in proteome) and its 'Context' is the short flank sequences surrounds this 'Object' (Fig. 1), 'Context' determines a unique 'Object' in a genome and 'Objects' have identical 'Contexts' in two genomes are thought to be homologous. We used an unified formula '$C_{a1,a2,a3..}O_{b1,b2,b3..}$' to donate the concrete 'Context-Object' structure (definition refer to Fig. 1) used for genomes indexing; for computational efficiency, '$C_{a1,a2,a3..}O_{b1,b2,b3..}$' are set to fixed length through genomes indexing, for example, in this study, structure '$C_{K,K}O_1$' we used have a 'Context' length of 2K. 'Context-Object' of different structures yield very close result tree in principle, given the 'Context' length is long enough to avoid it randomly appears in the genome. The probability (p) of the 'Context' of '$C_{K,K}O_1$' randomly appears in a genome of length G is given by the equation: $p = G/4**2K$ assuming the genome is a random sequence, for a normal prokaryotic genome of 5M length, this equation yield K>=8, given $p < 0.01$.

For each genome, the program would build an index table of the genome in which each 'Object' of the genome is indexed by its 'Contexts', once all genomes are indexed, for each pair of genomes, the program then counts the total number of shared 'Contexts' (N) and the number of shared 'Contexts' with non equal 'Objects' and their ratio calculated as the distance between the pair of genomes (Fig. 1). Once all pair wise distances are calculated, a distance matrix would be generated and neighbor-joining (NJ) method (10) are applied to taking the distance matrix to construct phylogenetic trees. If a 'Context' appears repeatedly in the genome sequence, the 'Context' wouldn't be taking into account unless all its 'Objects' are identical. For unassembled raw sequencing data of organisms, '$C_{K,K}O_1$' are derived from reads instead, and the filter policies (see below) would be applied to filter '$C_{K,K}O_1$' with low quality values.

2. Low quality value tuple filtering policies

In the fastq format NGS datasets, the quality value of a nucleotide is coded as a single symbol using ASCII code. For every tuple of length 2K+1 (yield a '$C_{K,K}O_1$') in read, we obtained its corresponding quality value substring and decoded it using the formula: Quality value = ord(ASCII code symbol) – 64 (SRA data from NCBI has Quality value = ord(ASCII code symbol) – 33). The tuple's quality value was represented using the minimal quality value existing in the quality value substring. Any tuple with a quality value lower than 10 was filtered, for tuples with a quality value greater than 10, we count their accumulated appearance times in the NGS dataset of an organism and those tuples appeared only one time in the dataset were filtered.

3. NGS data Simulation

Perfect NGS data was generated using in house Perl script; the simulated reads were 75-bp long substrings taken randomly from genome in either positive or negative strands. FastQ format sequencing data was generated using the tool 'Maq simulation' in MAQ package with these arguments: Using haploid model; set 'rate of mutations' and 'fraction of 1bp indels' to 0. The package 'CO-phylum' is available following this link https://secure.filesanywhere.com/fs/v.aspx?v=8a6a6687596775af9caa#. All accession numbers used in this study are available in Table S1.

Results

1. Test CO-phylum by '$C_{K,K}O_1$' of varying K

We first tested CO-phylum ('$C_{K,K}O_1$') by construct the tree of 13 organisms under genus *Brucella* and compared the tree with alignment-based method. The phylogenetic relationships of genus *Brucella* have been previously carefully explored by Foster *et al*., based on genome-wide orthologous SNPs alignments (12). They generated the same phylogenetic relationships using NJ, Maximum parsimony (MP) and Maximum likelihood methods; thereby it is rational to utilize their phylogenetics tree as a reference. The comparisons indicated CO-phylum trees with '$C_{K,K}O_1$' of varying K are all highly close to the reference one in both topology and branch length (Fig. 2), this illustrate CO-phylum is quite robustness with the choice of K.

2. Compare the accuracy of CO-phylum tree with that of other alignment-free methods

We then applied CO-phylum to explore the phylogeny of 26 strains of *Escherichia coli/Shigella*, using CO-structure '$C_{12,12}O_1$'. The reference tree came from the work of Zhemin Zhou *et al*. in which they concatenated the alignments of the 2034 core genes of *Escherichia coli/Shigella* (13) and the maximum likelihood (ML) method was employed to infer the phylogenetic relationships. CO-phylum again showed highly similar topology and branch length to the reference tree. This 26 *Escherichia/Shigella* phyloneny are also constructed using <u>CVtree method</u> and <u>Kr method</u> (14) (Fig. 3). The most significant difference of CVtree's result with that of other methods is that the genus *Shigella* violates the monophyleticity of the genus *Escherichia* but not the monophyleticity of the *E. coli* strains, such a result are also achieved by FFPs method (11). Noting CVtree and FFPs are similar methods based on genome 'word frequency' feature while CO-pylum and alignment-based method stand on the basis of molecular evolution theories, a possible explanation would be that CVtree and FFPs represent the taxonomy based on genome feature while CO-pylum and alignment-based method represent the phylogenetic relationship. In addition, unlike CO-pylum, the branch length in CVtree is not proportion well with the reference tree. Another alignment-free method Kr is developed for efficient estimation of pair wise distances between genomes and "more accurate than model-free approaches including the average common substring" (14). However, according to the test based on our data, Kr tree are still much less accurate than CO-pylum by both topology and branch length (Fig. 3), though the branch length seems much more rational than CVtree method, this illustrate how convince and accurate CO-phylum is in close related organisms phylogeny. The only inconsistence between CO-phylum tree and reference tree lie in the branch of *E. coli CFT073/E. coli 536/E. coli ED1a* which also observed in other two alignment-free methods, we found the inconsistence of this branch could been avoid by deleting only *E. coli CFT073* then regenerated the tree (data not show). It seems that the accuracy of CO-phylum as well as CVtree and Kr methods might be affect slightly if genomes undergo extensive reorganization (duplication, recombination etc.) like *E. coli CFT073*(15), in fact, this is a common problem for all distance based tree building methods, for they just can't discriminated which genome segment had gone through reorganization, might the only way to avoid the problem is multiple genome alignment.

3. Construction of phylogenetic relationships using simulated and real NGS data

We then test CO-phylum utilizing the simulated 75-bp Illumina NGS data. Other NGS sequencing platforms might generate reads of different lengths, but our method would not be affected by read length in principle. In the first stage, we did not consider the quality value of the nucleotide base, and the sequencing was supposed to be perfect (no sequencing error). We only explored the relationships between sequencing depth and the performance of CO-phylum. Simulated sequences datasets of varying sequencing depth (Coverage = 2X, 6X, 16X, 30X, 50X) were generated for each organism of *Brucella*. For simulated sequences datasets of each coverage, we constructed a *Brucella* phylogenetic tree by CO-phylum using '$C_{15,15}O_1$' and the '$C_{15,15}O_1$' completed genome tree used as a reference. Comparison showed sequences datasets of all coverage generated trees with identical topology except that of coverage 2X (Fig. S1). Our analysis showed the 'Contexts' count from the 2X coverage datasets were only 73.67% of 'Contexts' counts from the complete genome which is far less than those of the high coverage datasets (Table S2), it indicate too low depth of sequencing lead to inadequate 'Contexts' and affect the accuracy of distance calculated thereby might generate a biased tree. Hereafter, we test CO-phylum utilizing simulated fastQ format sequencing data, namely, NGS data with sequencing quality value which was generated by the tool 'Maq simulation' in MAQ package. When processing with NGS data CO-phylum applied a filter policy to filter tuples with quality values under certain cut-off would be filtered (see Methods for details). Before constructing the phylogenetic trees, the proportions of tuples pass through the filter policy were analyzed according to varying sequencing depth (Table 2). It showed our filter policies could capture most tuples of the genome when the sequencing depth was merely higher than l6X which are overpass by most prokaryotic NGS projects; we thus constructed phylogenetic trees using simulated NGS data with those sequencing depth over 16X and compared it with the tree constructed using complete genome (reference tree). Comparisons indicated all simulated NGS data trees have identical topologies and close branch length with the reference one (Fig. S2). We also considered that, in practice, CO-phylum will probably deal with NGS data from various independent sequencing projects, and thus we also constructed a 'mixed coverage' tree in which different organisms used sequencing data of different coverage (Table S3), the 'mixed coverage' tree again gave identical topology and close branch length with reference one (Fig. S2).

Next we test CO-phylum method utilizing the real NGS data of 29 *Escherichia coli* organisms that downloaded from NCBI Short Reads Archive (SRA) database; this data set already contain all the *Escherichia coli* organisms whose NGS raw data and assembled genomes are both available in NCBI. Given the accuracy of the CO-plum tree constructed based on assembled genomes has been proved in previous section, it is rational to take the assembled genomes tree as a reference. Comparisons indicated NGS raw data tree was almost identical with the assembled genome tree (Fig. 4), noting that these NGS raw data come from all three popular sequencing platforms: 454, illumina and SOLID (Table S1), CO-phylum would be quite robustness to choice of sequencing platforms and it illustrate without doubt that CO-phylum could be applied in the phylogenetic analysis of unassembled NGS data and achieved quite convince results.

4. Distance analysis and comparison

Beyond the comparison of phylogenetic relationships, we then go further to explore how close the CO-plum distance (CO-distance) to p-distance is, given an alignment is known. The 40 way *Escherichia/Shigella* genomes alignment (including all 26 *Escherichia/Shigella* genomes taken for phylogenetic analysis above (Fig. 3)), which was produced by multiple-alignment tool progressiveMauve (16), was utilized to calculate pair wise genome p-distances (gaps were completely deleted). Pair wise distances are also computed using CO-phylum, CVtree and Kr methods respectively, then linear regression are performed between p-distance and each of the there distance to evaluated how well they fit with p-distance. All 325 pair wise distances produced among the 26 *Escherichia/Shigella* genomes were plotted (Fig. 5), it showed that the CO-distance fit astonishing well with p-distance, almost all the points stay in a line, this yield a linear model: CO-distance = 0.70p-distance - 0.00038 with a correlation coefficient of 0.9919 for '$C_{9,9}O_1$' and CO-distance = 0.66p-distance - 0.00045 with a correlation coefficient of 0.9907 for '$C_{12,12}O_1$', As a comparison CVtree-distance and Kr-distance yield linear model：CVtree-distance = 4.27p-distance + 0.081 with a correlation coefficient of 0.3464 and Kr-distance = 0.85p-distance - 0.003 with a correlation coefficient of 0.7796，respectively. It is no doubt that CO-distance significantly outperforms CVtree-distance and Kr-distance. The significant linear relationship between CO-distance and p-distance also are also kept on other close related organisms based on out test data of primate Mitochondria DNA alignment (data not shown). This strong linear relationship explained why the CO-phylum tree agree so well with the alignment-based tree and illustrate that the CO-phylum is quite convince and accurate in the application of phylogenetc analysis of closed related organisms.

5. Performance on higher taxonomy level

We also test CO-phylum on higher taxonomy level, totally 63 organisms from 18 genus under *Enterobacteriaceae* are used to construct the phylogenomics tree and compare with the tree build based on 16S rRNA sequences from the same organisms through Tree Builder on the Ribosomal Database Project website which create a phylogenetic tree using the Weighbor weighted neighbor-joining algorithm (17). It shows 16S rRNA tree and CO-phylum tree agree very well in general, however, genus *Enterobacter* and *Citrobacter* are polyphyletic group and genus *Yersinia* is paraphyletic group in 16S rRNA tree while all these genus formed single clade in CO-phylum tree, illustrating CO-phylum is still more accurate and also much higher resolution than 16S rRNA tree on family level (Fig. 6). We then tested CO-phylum on even higher taxonomy level, the accuracy of CO-phylum didn't significantly diminish until go up to class level. We build tree based on 70 genomes from 14 order under class *Gammaproteobacteria*, it indicated *Enterobacteriales*, *Xanthomonadales*, *Pasteurellales* and *Thiotrichales* are still forms a clade generally while other orders formed paraphyletic or polyphyletic group in the tree (Fig. S3). Its performance on taxonomy levels higher than class were also test but turn to be even worse than on class level (data not show).

Discussion

CO-phylum was initially designed for unassembled NGS data phylogenetic analysis，still there isn't any other assembly-free phylogenetics method so far, in this respect, CO-phylum is unique. Furthermore CO-phylum tree have ultrahigh resolution because whole genome information are taken advantage to construct phylogenetec relationship and CO-phylum produces highly accurate phylogenetic relationship given organisms analyzed are close related, say, under the same family, it yields distance almost linear with p-distance, this explain why CO-phylum tree was so close to the tree constructed by alignment-based methods. As a contrary, the distance of other alignment-free methods like CVtree, FFPs and information-based methods are not based on theory of molecular evolution principally thereby exhibit no explicit relationship with p-distance. The superior of these methods lie in high level taxonomy, in fact some of them like FFPs and information based methods even could used at literature classifying (18).

The data structure employed and distance generate procedure in CO-phylum are similar with that of CVtree and they have equal algorithm efficiency, both of them takes $O(L)$ steps to compute the distance between two genomes of length L and takes $\binom{N}{2}$ steps compute all pair wise distances of N organisms, but they are essentially different method: CO-phylum index 'Object' by 'Context' while CVtree use 'K-mer' to index its normalized frequencies which got by subtracting (K-1)-mer frequencies from whole proteome K-mer frequencies (5), that is, the genome must be assembled for the applicability of CVtree (this also the case for other phylogenomics approaches developed so far), in addition, for each K-mer, CVtree must compute its whole proteome frequencies, these exclude the applicability of CVtree and other phylogenomics approaches to unassembled NGS data, furthermore, differential of 'normalized K-mer frequencies' in CVtree don't have any explicit biological meaning while a non equal 'Object' could represent a SNP, this explains why CO-phylum is much more accurate than

CVtree and alike methods.

## Acknowledgments

Thanks Z. Hongxiang for helpful discussion. The project was supported by the National Natural Science Foundation of China (NSFC) (30890034 and 31000552).

Figure Legends:

Fig. 1 CO-phylum algorithm overview

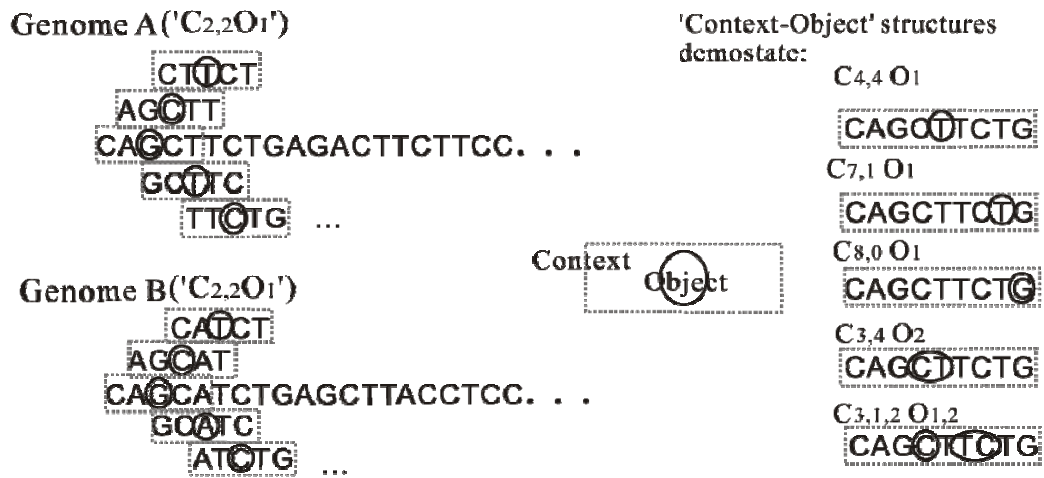

| Contexts A | Objects A | Contexts B | Objects B |
|---|---|---|---|
| CTCT | T | CACT | T |
| AGTT | C | AGAT | C |
| CACT | G | CACA | G |
| GCTC | T | GCTC | A |
| TTTG | C | ATTG | C |
| .... | . | .... | . |

| i | Shared Cont AB | ObjA | ObjB |
|---|---|---|---|
| 1 | GCTC | T | A |
| 2 | CACT | G | T |
| 3 | ATTG | C | C |
| . | ... | . | . |
| . | ... | . | . |
| N | ... | . | . |

$$\text{for } i = 1..N, \quad I = \begin{cases} 0 \text{ if ObjA} = \text{ObjB} \\ 1 \text{ if ObjA} \neq \text{ObjB} \end{cases}$$

$$\text{Distance}(A,B) = \frac{\sum_{i=1}^{N} I}{N}$$

Fig. 2 *Brucella* phylogeny by CO-phylum ($C_{K,K}O_1$)

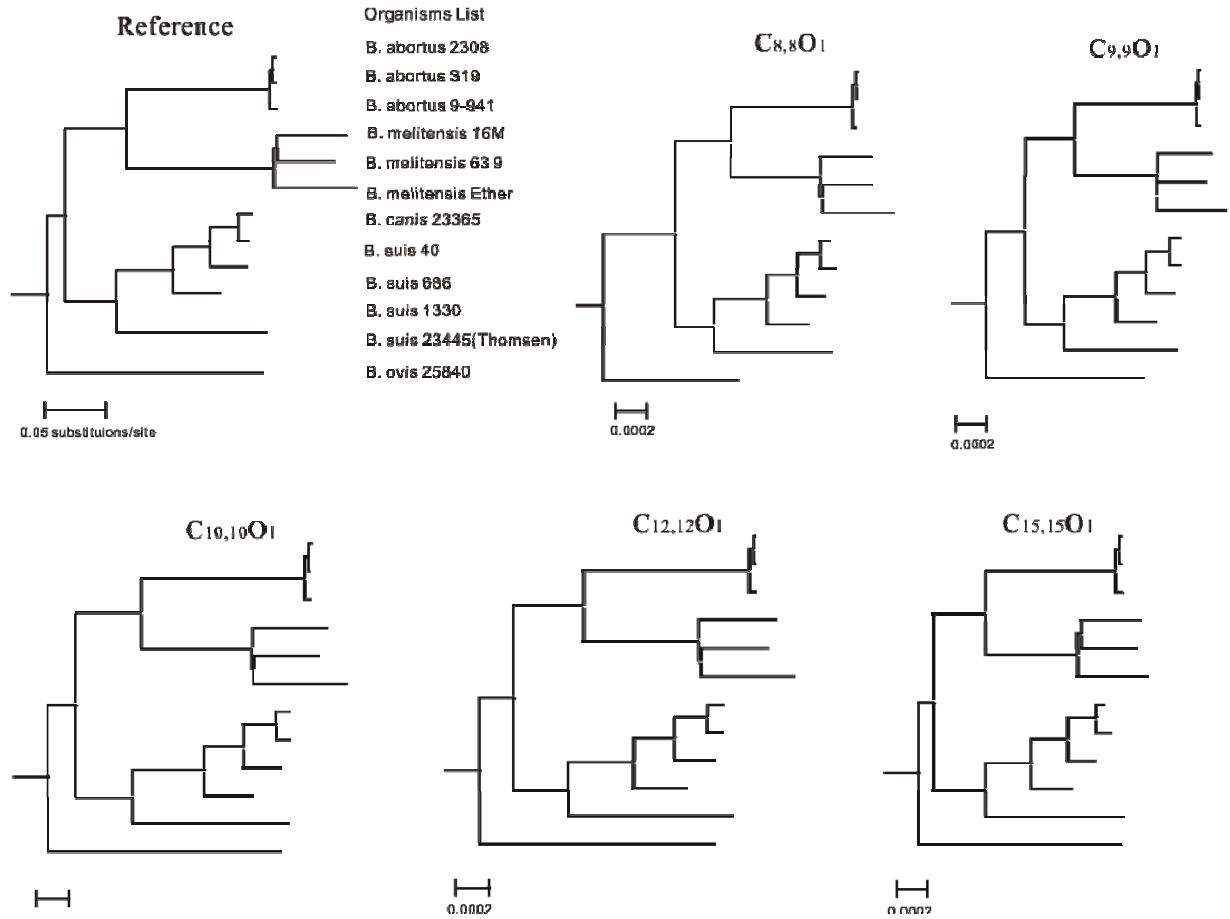

Use *Ochrobactrum anthropi* as the outgroup taxa. All the trees have same organisms order list.

Fig. 3 26 *Escherichia coli/Shigella* strains phylogeny by four different methods

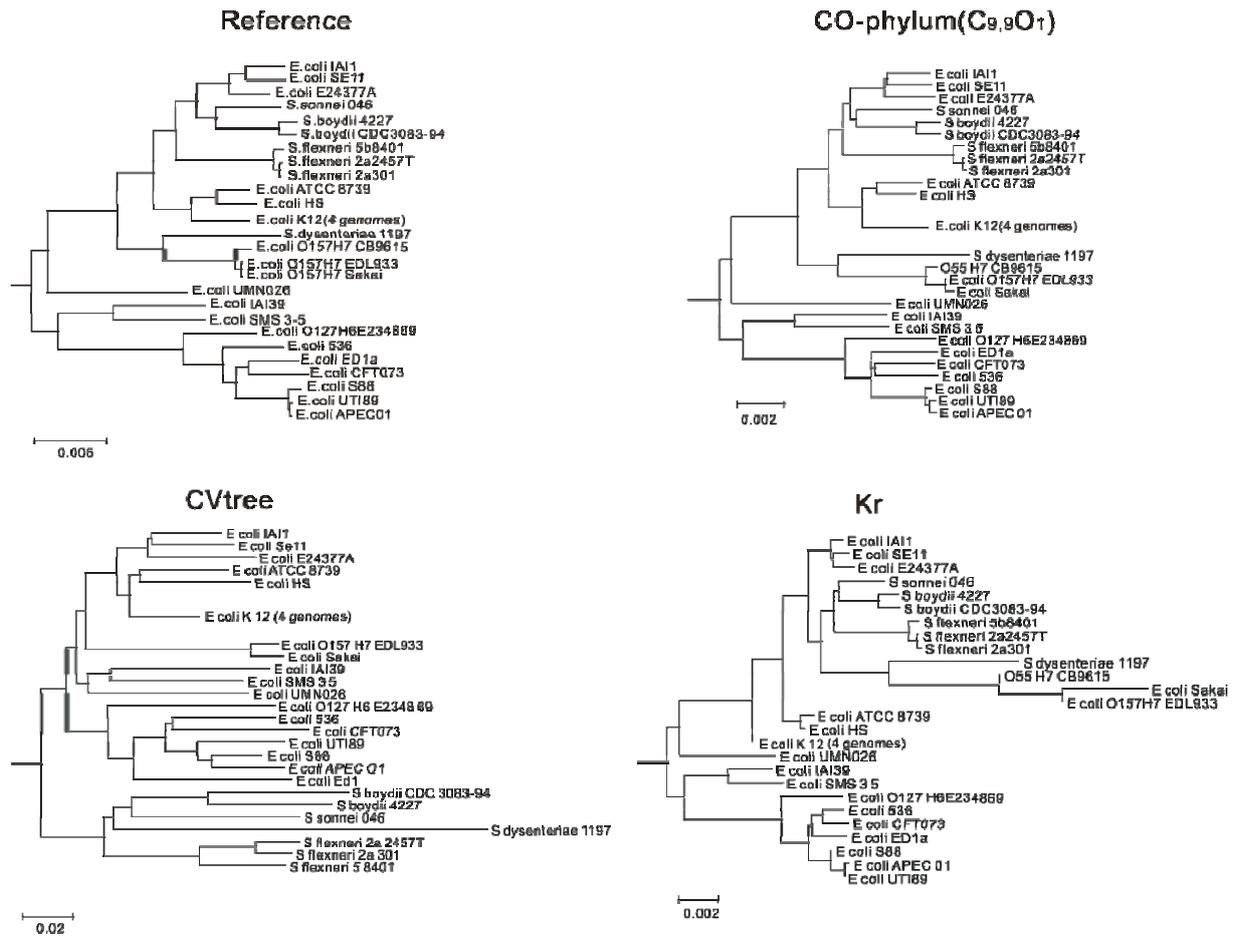

Use *Escherichia fergusonii* as the outgroup taxa

Fig. 4 Test CO-phylum using real NGS data

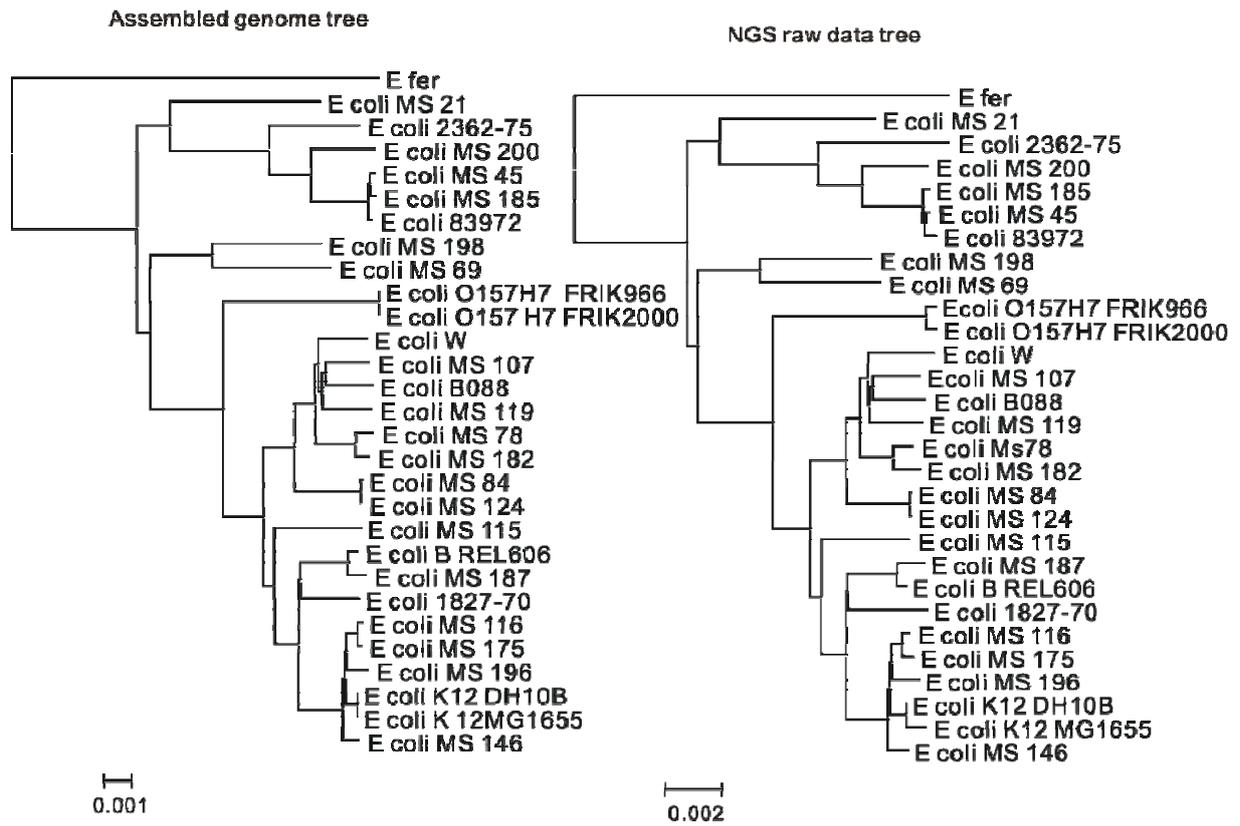

Fig. 5 Linear regression with p-distance using distances generated by different methods

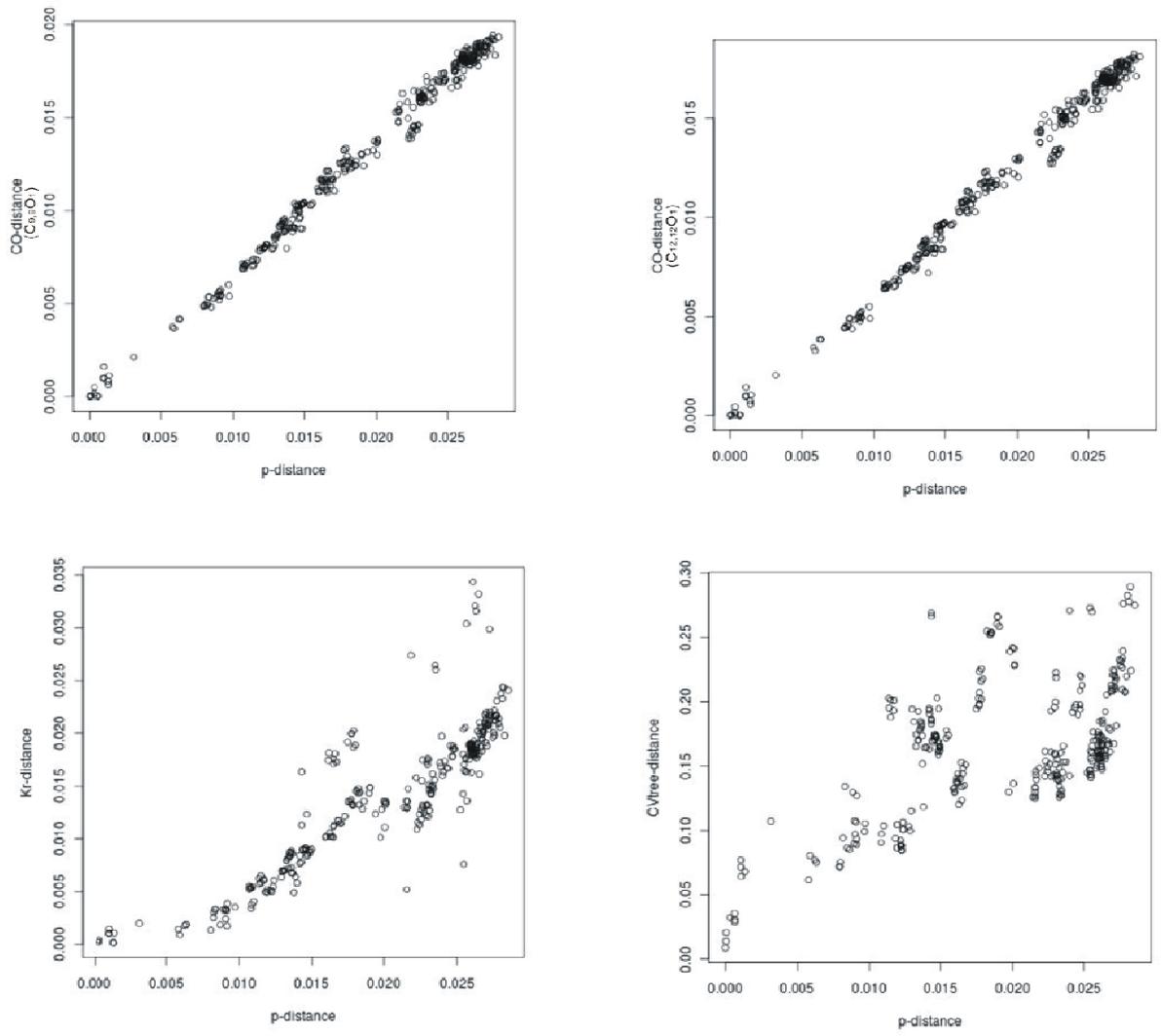

Fig. 6 Test the performance of CO-phylum on family level

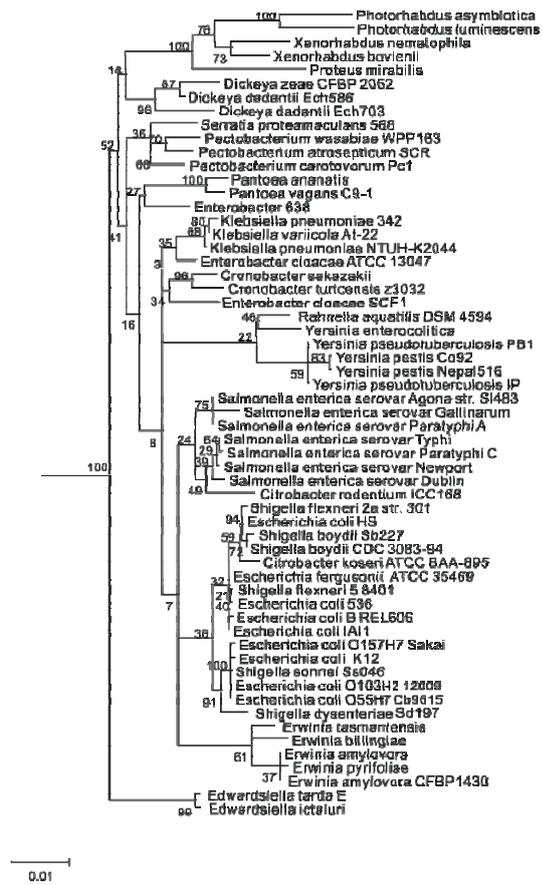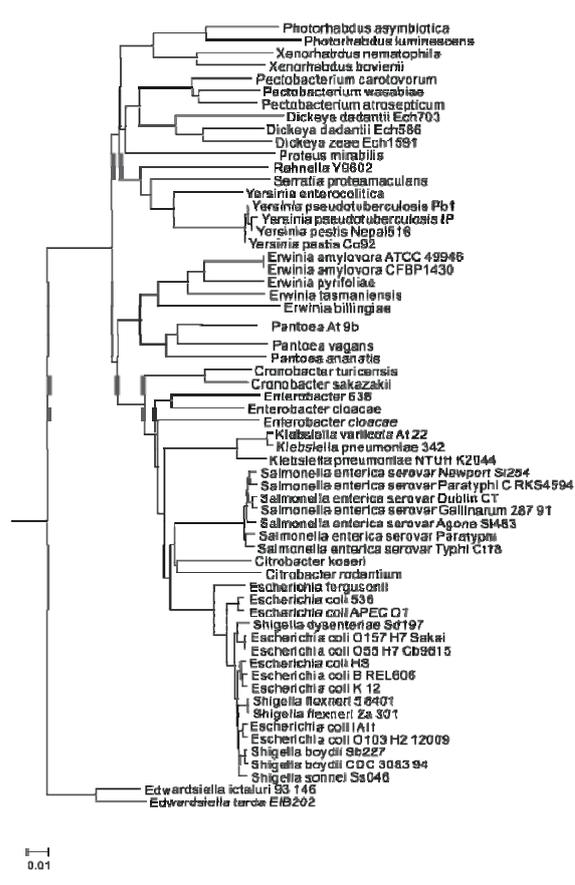

Use *Pasteurella multocida* as the outgroup taxa.